\documentclass[fleqn,usenatbib]{mnras}

\usepackage{txfonts}

\usepackage{graphicx}	% Including figure files
\usepackage{amssymb}	% Extra maths symbols
\def\app#1#2{%
  \mathrel{%
    \setbox0=\hbox{$#1\sim$}%
    \setbox2=\hbox{%
      \rlap{\hbox{$#1\propto$}}%
      \lower1.1\ht0\box0%
    }%
    \raise0.25\ht2\box2%
  }%
}
\def\approxprop{\mathpalette\app\relax}

\title[EM counterparts of 3D GRMHD jets]{EM counterparts of structured jets from 3D GRMHD simulations}

\author[Kathirgamaraju, Tchekhovskoy, Giannios \& Barniol Duran]{Adithan Kathirgamaraju$^{1}$\thanks{E-mail: akathirg@purdue.edu (AK)}, Alexander Tchekhovskoy$^{2}$, Dimitrios Giannios$^{1}$, 
and \newauthor Rodolfo Barniol Duran$^{3}$
\\
$^{1}$ Department of Physics and Astronomy, Purdue University, 525 Northwestern Avenue, West Lafayette, IN 47907-2036, USA\\
$^{2}$Center for Interdisciplinary Exploration \& Research in Astrophysics (CIERA),
Physics \& Astronomy, Northwestern University, Evanston, IL 60208, USA\\
$^{3}$Department of Physics and Astronomy, California State University, Sacramento, 6000 J Street, Sacramento,  CA 95819-6041, USA 
}

\date{Accepted XXX. Received YYY; in original form ZZZ}

\pubyear{2015}

\hypersetup{draft}
\begin{document}
\label{firstpage}
\pagerange{\pageref{firstpage}--\pageref{lastpage}}
\maketitle

\begin{abstract}
GW170817/GRB170817A has offered  unprecedented insight into binary neutron star post-merger systems. Its Prompt and afterglow emission imply the presence of a tightly collimated relativistic jet with a smooth transverse structure. However, it remains unclear whether and how the central engine can produce such structured jets. Here, we utilize 3D GRMHD simulations starting with a black hole surrounded by a magnetized torus with properties typically expected of a post-merger system. We follow the jet, as it is self-consistently launched, from the scale of the compact object out to more than 3 orders of magnitude in distance. We find that this naturally results in a structured jet, which is collimated by the disk wind into a half-opening angle of roughly $10^\circ$, its emission can explain features of both the prompt and afterglow emission of GRB170817A for a $ 30^{\circ}$ observing angle. Our work is the first to compute the afterglow, in the context of a binary merger, from a relativistic magnetized jet self-consistently generated by an accreting black hole, with the jet's transverse structure determined by the accretion physics and not prescribed at any point.

\end{abstract}

% Select between one and six entries from 

%%%%%%%%%%%%%%%%%%%%%%%%%%%%%%%%%%%%%%%%%%%%%%%%%%

%%%%%%%%%%%%%%%%% BODY OF PAPER %%%%%%%%%%%%%%%%%%

\section{Introduction}
The conjoint detection of the first binary neutron star (NS) merger, GW170817, in both gravitational and electromagnetic waves heralds a new era in multi-messenger astronomy \citep{abbott2017,abbott2017b}. One of the electromagnetic (EM) counterparts associated with this event was a burst of gamma rays detected about 1.7 seconds after the merger and lasted for $\sim 0.6$ seconds (GRB 170817A) (e.g., \citealt{goldstein2017,savchenko2017}). This detection provides the most conclusive evidence yet that binary NS mergers are indeed a progenitor of short gamma-ray bursts (GRBs) as hypothesized a few decades ago \citep{blinnikov1984,paczynski1986,eichler1989,narayan1992}. However, this short GRB and its associated afterglow emission exhibit some peculiar characteristics that are unlike any other burst. For example, it was a faint short GRB despite being, by far, the closest detected to date, and its afterglow showed a shallow rise for several months as opposed to typical afterglows that show a decline from the beginning
(e.g., \citealt{fong2017,margutti2017,troja2017}).

In standard GRB theory, the EM emission is produced by a highly relativistic jet launched by a compact object, either a NS or a black hole (BH). In the case of short GRBs, this jet is believed to be produced by the remnant of a binary NS or NS-BH merger (for a review see e.g., \citealt{lee2007,nakar2007,metzger2012}). The brief flash of high energy X-rays and gamma rays typically lasting less than 2 seconds (the `prompt' emission associated with the sGRB) is attributed to an internal mechanism within the jet that is not yet well understood. As the jet propagates through the external, interstellar medium (ISM), it drives a shock that sweeps up and accelerates external particles which in turn radiate predominantly via synchrotron emission (the GRB `afterglow') that can last up to several months.
The most widely used jet model when calculating the emission from a GRB jet is a `top-hat' jet, where the jet is a conical outflow and the properties of the jet within this cone (e.g., Lorentz factor and energy) are assumed to be constant. Beyond this cone, the jetted outflow ceases abruptly. In the past, the top-hat jet has been able to reproduce the observed characteristics of GRBs, but it fails to explain both the prompt and afterglow emission of GW170817 (e.g., \citealt{granot2017}). The key difference is that due to the gravitational wave trigger and impressive follow-up effort of GW170817, it might be the only known GRB viewed at an angle larger than the jet's core (i.e. ``off-axis"), for which the prompt and afterglow emission has been detected. The emission received by off-axis observers can vary greatly depending on how the jet's power and Lorentz factor vary as a function of polar angle. Previous studies have suggested that a more realistic model is that of a structured jet, where the Lorentz factor and energy flux vary smoothly within the jet as a function of polar angle (e.g., \citealt{rossi2002,kumar2003,Aloy05,janka2006}). Recent works have used structured jet models to investigate the characteristics and feasibility of detecting the prompt and afterglow emission from such jets as possible EM counterparts to GW events \citep{lamb2017,lazzati2017,kathir2017}. Now, a year after the detection of GW170817, the structured jet model has been able to successfully reproduce the observed afterglow and can explain some of the peculiar characteristics related to the prompt emission of the short GRB, leading to the interpretation that GRB 170817A may have been a regular short GRB but viewed off-axis (e.g., \citealt{alexander2018,davanzo2018,dobie2018,lamb2018,lazzati2018,margutti2018,resmi2018,xie2018,troja2018b}). The detection of superluminal motion in the outflow also provides strong observational evidence for the structured jet model (\citealt{mooley2018}). 

How the jet distributes its power as a function of the polar angle and distance from the central engine remains an open question. Most numerical studies initiate jet simulations in post merger systems by injecting the jet at a length-scale of a few orders of magnitude larger than that of the central engine. In particular, top-hat jets are injected into an ambient gas to follow its hydrodynamic interactions as it breaks away from the confining medium. Sufficiently far from the break out scale, the jet turns conical and its structure can be determined. A more realistic investigation into jet structures must begin at the central engine, taking into account the initial conditions of the compact object and its surroundings. It must include a consistent jet launching mechanism from the compact object and follow the jet as it collimates and accelerates out to large distances. In this work, we use 3D general relativistic magnetohydrodynamic (GRMHD) simulations to study the jet structure with many of the factors above taken into account. We start with a black hole torus system with properties typically expected from a binary NS post-merger system, and follow the MHD-driven jet, launched self consistently via the accretion and rotation of the compact object, as it is initially collimated by the surrounding disk winds, up to a point where these interactions with the disk winds become insignificant, after which the structure of the jet is extracted. Building upon our previous work \citep{kathir2017}, we calculate the emission profiles (prompt and afterglow) produced by the structured jet from these improved simulations.

In Section \ref{setup} we describe the simulation setup. In Section \ref{results} we present the simulation results, which include the jet structure, its emission profile and comparison with the latest observations of GRB 170817A. We discuss these results and conclude in Section \ref{conclusions}.

\section{Numerical setup}
\label{setup}

\begin{figure}
	
\includegraphics[width=\columnwidth]{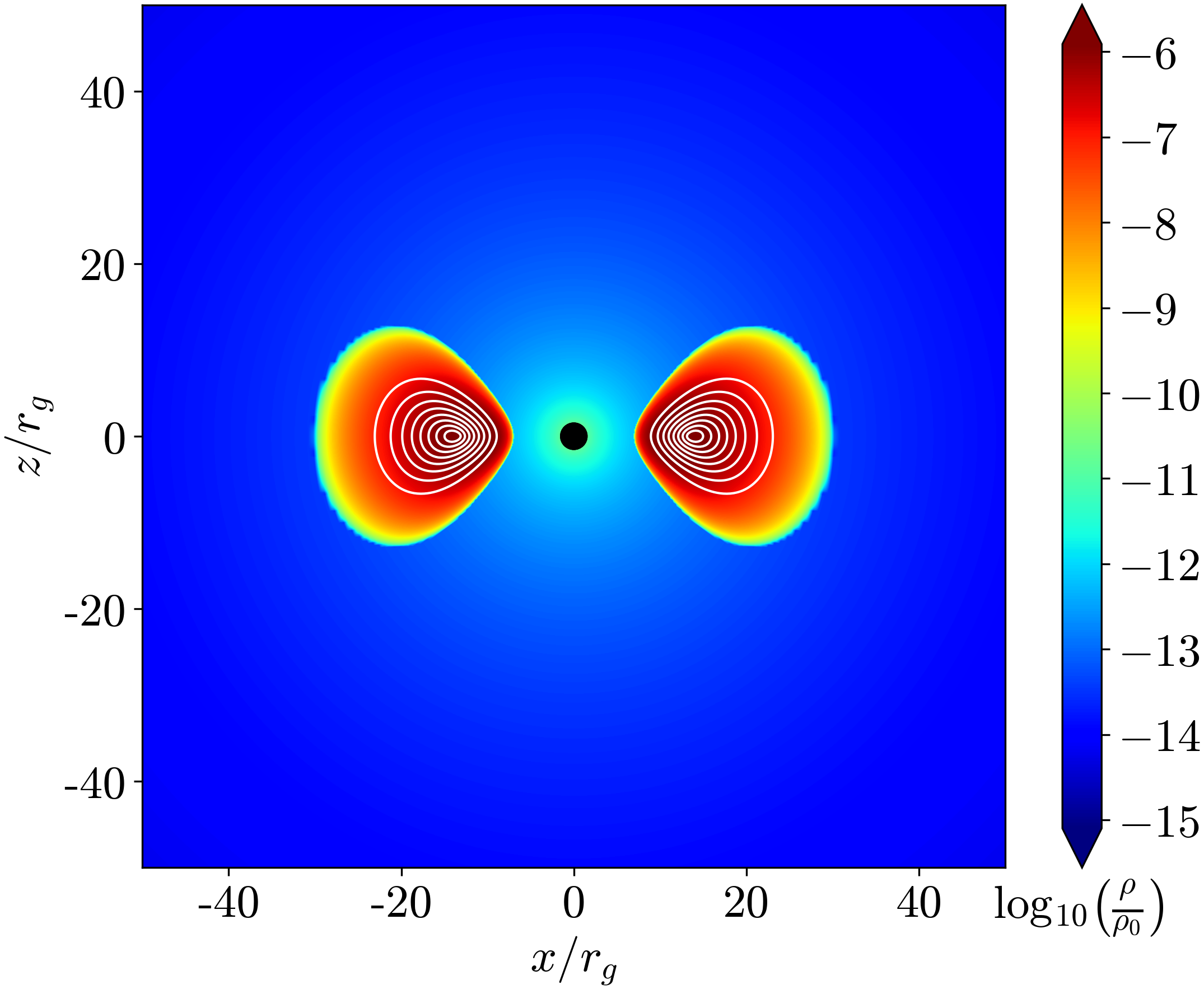}\\
\includegraphics[width=\columnwidth]{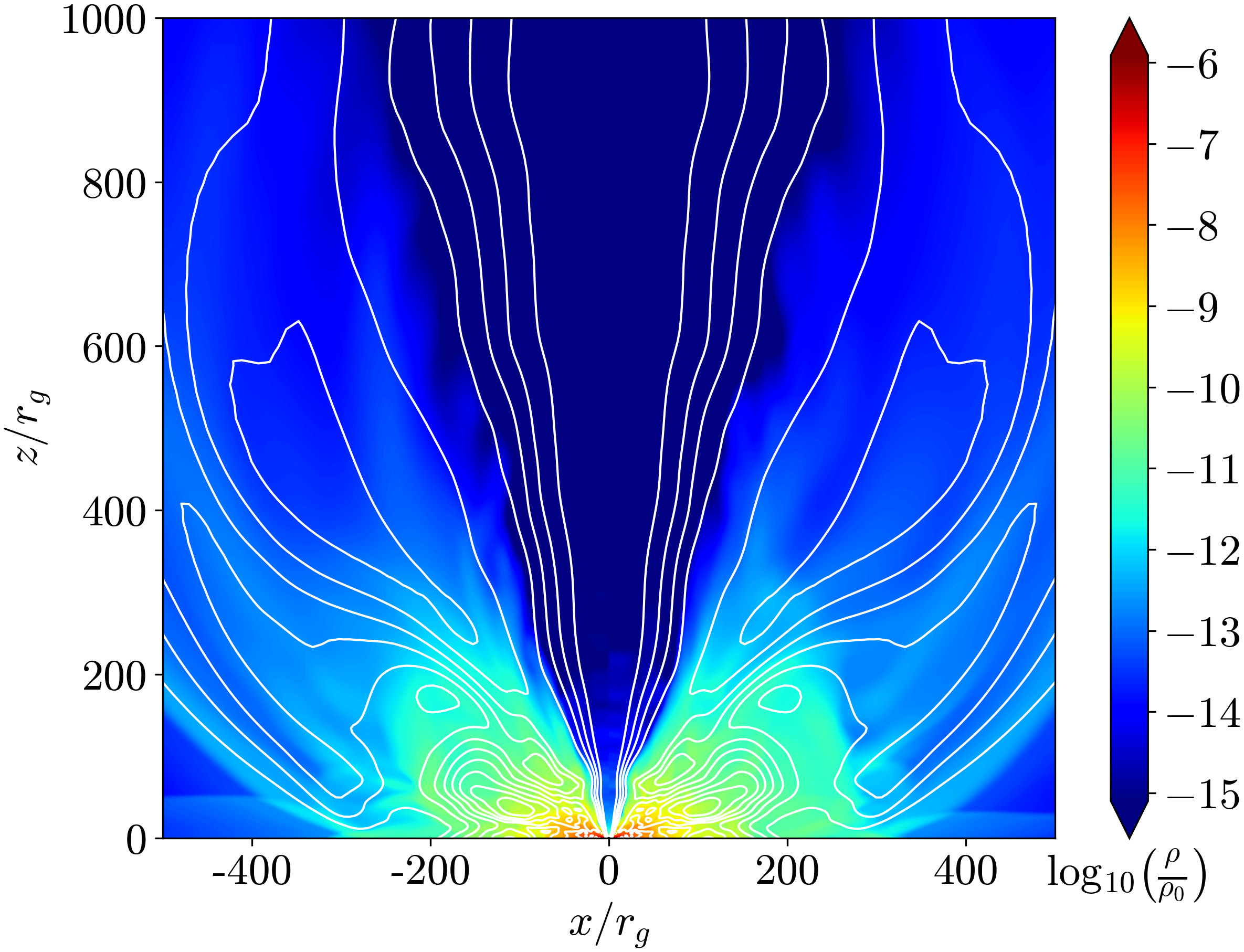}
\vspace*{-5mm}\caption{2D plots (vertical slice) of density and magnetic field line contours of the initial setup (top) and at $\sim 0.05$ seconds (bottom), axes are in units of r$_{\rm g}$ (the gravitational radius) and $\rho_0\approx 7\times 10^{16}$ g cm$^{-3}$. The compact object is at the origin, and the bottom panel shows only one of the two jets. The jet is initially collimated by the disk winds and eventually breaks out at $z\sim 1000$ r$_{\rm g}$, after which it propagates conically.}
    \label{initialsetup}
\end{figure}

A detailed description of the simulation setup can be found in \citealt{fernandez2018} (their model B3d\footnote{ These simulations were run before GW170817 was detected, thus it does not use the properties inferred from observations of this event.}) and will be summarized here. Simulations are carried out using HARMPI\footnote{Available at
  https://github.com/atchekho/harmpi \label{fn:harmpi}}, an enhanced version of the serial open-source 
code HARM \citep{gammie_2003,noble_2006} with modifications that consider additional physical processes such as neutrino cooling and nuclear recombination. The initial setup consists of 3 solar mass (M$_{\odot}$) BH  wih spin 0.8, surrounded by a torus of $\sim$ 0.03 M$_{\odot}$ embedded with a poloidal magnetic field prescribed by the vector potential $A_\phi\propto r^{5}\rho^2$ and having a maximum field strength of $4\times 10^{14}$ G, where $r$ is the radius in spherical coordinates and $\rho$ is density. The top panel of Fig. \ref{initialsetup} shows a contour plot of the density and magnetic field of the initial setup. Regions outside the BH and torus are set to the floor density, which initially drops off with radius as $\propto r^{-2}$, therefore density contributions from the post-merger and dynamical ejecta are not considered in this setup (e.g., \citealt{hotokezaka2013,sekiguchi2016}). Following the onset of accretion, a jet is launched $\sim 10^{-2}$ ms after the start of the simulation and $\sim 90\%$ of the jet (in energy) is ejected within the first $\sim 1.5$ seconds, which is consistent with the duration of the prompt phase of GRB170817A. The bottom panel of Fig. \ref{initialsetup} shows a contour plot of the density and magnetic field at $\sim 50$ ms. The polar, under-dense region consists of the jet which is surrounded by the denser disk winds that initially collimate the jet. By the time the jet reaches a length-scale of $\sim 1000$ r$_{\rm g}$ (r$_{\rm g}=GM/c^2$ is the gravitational radius of a BH with mass $M$), it propagates out of the confining winds and becomes conical, travelling radially outwards. The simulation is able to accurately track the jet starting from its launching region near the compact object, up to a distance of a few $\sim 1000$ r$_{\rm g}$.

\section{Results}
\label{results}

\subsection{Jet structure from simulations}
\label{structure}
\begin{figure}
\includegraphics[width=\columnwidth]{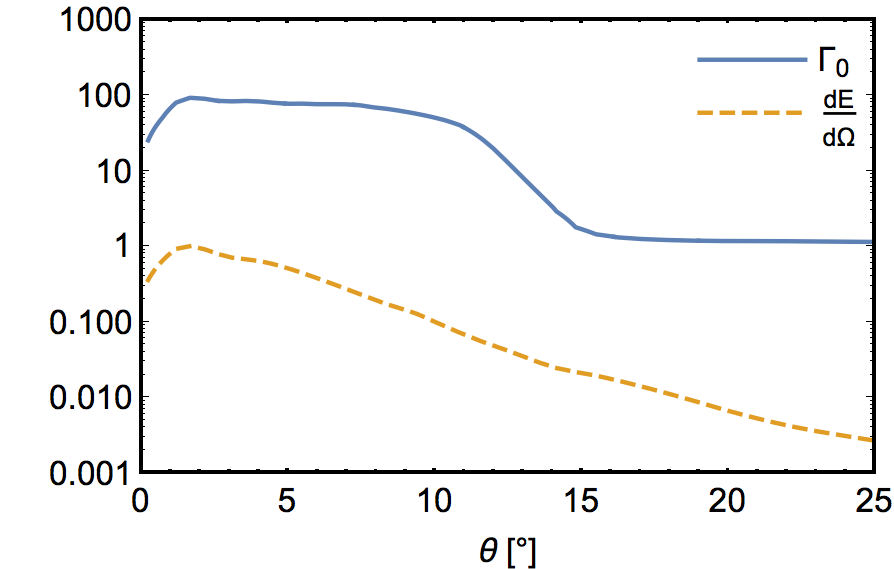}
\vspace*{-5mm}\caption{Jet structure self-consistently obtained from our post-merger remnant disk simulations showing the Lorentz factor of the jet $\Gamma_{0}$ (solid line, calculated as energy-flux-weighted ratio of energy flux to mass flux averaged over time, see Sec.\ref{structure}) and normalized jet energy per solid angle (dashed line). Both quantities have been averaged over azimuthal angle $\phi$ at a fixed radius of $\sim 2000$ r$_{\rm g}$.}
	\label{jetstructure}
\end{figure}

In order to calculate the observed emission from the jet we first need to extract its structure from the simulations. The required quantities are the Lorentz factor and energy per solid angle of the jet as a function of the polar angle $\theta$, the energy here includes the electromagnetic, thermal and kinetic energy (without rest mass). Current 3D GRMHD simulations are unable to follow the jet to the scales where the prompt emission and afterglow emission take place (beyond $\sim 10^{6}$ r$_{\rm g}$). We are therefore only able to extract the quantities up to the distance where the simulations are still accurate ($\sim 2000$ r$_{\rm g}$), and this structure is plotted in Fig. \ref{jetstructure}. The dashed line in Fig. \ref{jetstructure} shows the normalized energy per solid angle (dE/d$\Omega$) of the jet versus polar angle $\theta$, obtained by summing the energy flux over time at a fixed radius of $\sim 2000$  r$_{\rm g}$ and averaging over azimuthal angle $\phi$. The jet is dominated by electromagnetic and kinetic energy with a small thermal component. This distribution roughly follows a power law decline $\approxprop\theta^{-3.5}$ between $\sim 5- 15^{\circ}$, the total energy of one jet is $\sim 10^{51}$ erg. The next quantity we require is the Lorentz factor distribution of the jet, however the jet may not have undergone complete acceleration at the distances mentioned above. Therefore we 
find the ratio of the total energy flux to mass flux ($\mu$), which determines the maximum achievable Lorentz factor in MHD jets, and use $\mu$ as an estimate for the terminal Lorentz factor of the jet. We calculate the energy-flux--weighted $\mu$ averaged over time and $\phi$ at a fixed radius of $\sim 2000$ r$_{\rm g}$, as a reliable estimate for the terminal Lorentz factor ($\Gamma_{0}$) of the jet as follows
\begin{equation}
\Gamma_{0} = \frac{\int\mu\,{\rm T}^{r}_t\,d\phi\,dt}{\int{\rm T}^{r}_t\,d\phi\,dt},
\end{equation}
where $T^{r}_t$ is the component of the electromagnetic stress-energy tensor containing the radial energy flux (electromagnetic, thermal and kinetic energy with the rest mass energy subtracted). The solid line in Fig. \ref{jetstructure} shows the jet Lorentz factor as a function of $\theta$: it remains roughly constant at the value of $\sim 100$ up to $\sim 10^{\circ}$ and then declines rapidly as a power law  $\approxprop\theta^{-11}$ to $\sim 1$ at $15^{\circ}$. In the next subsections, we will use the jet structure in Fig. \ref{jetstructure} to calculate the observed emission in both the prompt and afterglow phases. 

\subsection{Prompt emission profile}
\label{prompt}
\begin{figure}
\includegraphics[width=\columnwidth]{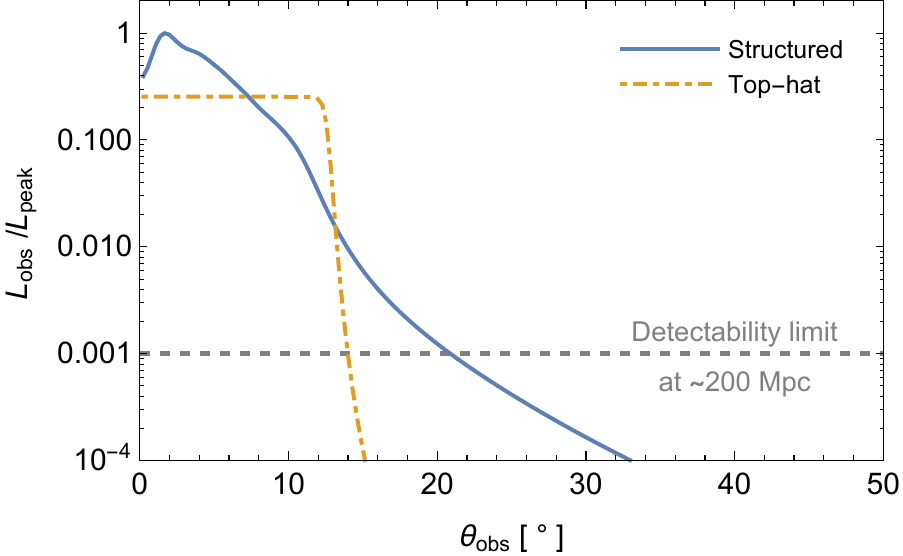}
\vspace*{-5mm}\caption{Normalized observed luminosity versus observing angle ($\theta_{\rm obs}$) for the prompt emission of structured (solid line) and top-hat (dashed line) jets. Structure of the jet is obtained from simulations (Fig. \ref{jetstructure}). The top-hat jet has half opening angle $13^{\circ}$ (the same angular extent taken for our structured jet -- see Sec. \ref{prompt}) and initial Lorentz factor 100 (equal to the Lorentz factor of the core of our structured jet). Horizontal dashed line indicates an estimate for a robust detection limit of a sGRB with an accompanying GW trigger by LIGO taking place at $\sim 200$ Mpc. The emission from the structured jet is detectable up to an observing angle of $\sim 20^{\circ}$ whereas the emission from top-hat jet falls much more steeply and is detectable up to $\sim 14^{\circ}$. For closer events like GW170817, this detectability limit can be $\sim 10$ times less (at $L_{\rm obs}/L_{\rm peak} \sim 10^{-4}$), allowing detection up to a viewing angle of $\sim 30^{\circ}$ in the case of a structured jet.}
    \label{promptprofile}
\end{figure}

Using the jet structure in section \ref{structure} we can calculate a luminosity profile for the prompt emission: the total prompt luminosity an observer can expect to receive versus observing angle. Following the calculations done in \citet{kathir2017}, we estimate the prompt emission luminosity as a function of the observing angle, $\theta_{\rm obs}$, which is the angle between the jet axis and the line of sight towards the observer. We assume a fixed fraction of the energy is radiated instantaneously and isotropically in the co-moving frame of the jet, and transform it to the observer frame (addressing dissipation mechanisms for the prompt emission is beyond the scope of this work). Fig. \ref{promptprofile} compares the prompt emission profile of observed luminosity $L_{\rm obs}$ (normalized to the peak luminosity $L_{\rm peak}$) versus observing angle $\theta_{\rm obs}$ for our simulated structured jet and a top-hat jet. We adopt a structured jet profile from the simulation, using the energy and Lorentz factor distributions in Fig.~\ref{jetstructure}, and only take into account material with Lorentz factor $\gtrsim 3$ (corresponding to $\theta\lesssim 13^{\circ}$) as implied by observations and constrained by the fact that slower components might initially be too optically thick to contribute to the prompt phase of GRB 170817A \citep{burgess2017,margutti2017}. The top-hat jet has half opening angle $13^{\circ}$ and a constant initial Lorentz factor of 100 which roughly correspond to the extent and maximum Lorentz factor of our structured jet, respectively. The total energy of the top-hat jet is set to be equal to that of the structured jet, thus enabling a fairer comparison. If the count rate of a typical sGRB is scaled to within the LIGO detectability distance ($\sim 200 $ Mpc at design sensitivity; \citealt{martynov2016}), it would be an extremely bright source of $\sim 10^6$ counts/s (e.g., \citealt{fong2015}), and since these typical short GRBs are viewed on axis, we can translate the normalized $L_{\rm obs}$ profile in Fig. \ref{promptprofile} to a count rate by scaling the peak of this profile  to $10^6$. The required count rate for a robust detection of short GRB that is coincident with a LIGO trigger is estimated to be $\sim 10^3$ counts/s \citep{connaughton2016}, which correspond to $L_{\rm obs}/L_{\rm peak} \sim 10^{-3}$. This limit is indicated by the horizontal dashed line in Fig. \ref{promptprofile}, and it gives us a robust detectability limit for a short GRB with an associated GW trigger from LIGO.
\subsection{Afterglow emission}
\label{afterglow}
The afterglow is calculated using the standard synchrotron emission techniques from forward shocks in an external medium of uniform density (e.g., \citealt{sari1998}). Jet spreading is taken into account following \cite{duffell2018b}, who utilize simulations to derive analytic expressions for the dynamics of a spreading jet. Although these expression were derived for a top-hat jet, they are still applicable to the core of our structured jet ($\lesssim 5^{\circ}$), where the energy and Lorentz factor do not change by more than a factor $\sim 2$. In order to calculate the afterglow we first need the initial structure of the blast wave, which is obtained from  the Lorentz factor and energy profile of our simulated jet in Fig. \ref{jetstructure}. The synchrotron emission is obtained semi-analytically by dividing up the blast wave into 10$^{4}$ patches (100 uniform segments along the $\theta$ and $\phi$ directions), calculating the synchrotron emission associated with the forward shock of each patch, and then summing the emission from all patches to obtain the total afterglow emission.
We assume each patch coasts at its initial Lorentz factor ($\Gamma_{0}$, shown in Fig.~\ref{jetstructure}) until the energy in the swept up, shocked medium is comparable to the initial energy of the patch, after which each patch decelerates as $\Gamma\beta \propto E^{1/2}R^{-3/2}$, where $E,R$ are the kinetic energy and spherical radius of the blast wave respectively and $\beta=\beta(\theta)$, $\Gamma=\Gamma(\theta)$ are the 3-velocity and Lorentz factor of the blast wave respectively
during the deceleration phase. The dependence of $\Gamma\beta$ on $R$ will steepen when the jet begins to spread and its observable implications will be discussed in Sec. \ref{conclusions}. We assume the synchrotron emission from each patch is radiated isotropically in the rest frame of the emitting region, and then transform this to the observer frame. The total afterglow emission is obtained by summing over all patches, covering the entire solid angle of the jet, and taking into account differences in the photon arrival time, $T_{\rm obs}=\int{\frac{dR}{\beta c}(1-\beta\,\cos\alpha)}$, where $T_{\rm obs}$ is the time in the observer frame, $\alpha$ is the angle between velocity vector of a patch of the jet and its line of sight towards the observer, $\beta$ is the velocity of the patch of the jet and $c$ is the speed of light. 

Figure~\ref{afterglow} shows afterglow light curves from our simulated structured jet and observed data points of GRB170817A afterglow for comparison. We neglect the counter-jet because its afterglow will be too faint to be detected. The parameters used to calculate the light curves in Fig. \ref{afterglow} are $E_{\rm j}= 5\times 10^{50}$ erg, $n\approx 0.05$ cm$^{-3}$, $\epsilon_{\rm e}\approx 0.01$, $\epsilon_{\rm B}=10^{-4}$, $p=2.17$, $\theta_{\rm obs}=30^{\circ}$, where  $E_{\rm j}$ is the true energy of the jet (without rest mass energy), $\epsilon_{\rm e}, \epsilon_{\rm B}$ are the fractions of the total energy in the shocked electrons and magnetic fields respectively, $n$ is the number density of the uniform external medium and $p$ is the power law slope of the distribution of shocked electrons. In reality, the value of $E_{\rm j}$ depends on the radiative efficiency of the prompt emission. From observations, this efficiency varies between a few percent to more than 90\% (\citealt{fong2015}), therefore we will assume a median value of 50\% efficiency, which means $E_{\rm j}\approx 5\times 10^{50}$ erg (half the value of the jet energy obtained from our simulations). The shallow rise in the afterglow (between $\sim 20-200$ days) occurs as the entirety of the jet becomes visible to off-axis observers. In our modelling, the slope of this rise (in a uniformly dense external medium) depends only upon the jet structure and observing angle. Since the jet structure is fixed from our simulations, we vary the observing angle and find that $\theta_{\rm obs} \approx 30^{\circ}$ produces a rise that matches the observations. The value of $p$ is inferred from observations (\citealt{margutti2018}) and rest of the parameters are adjusted to match the peak time and flux. These parameters are largely consistent with other works which model the afterglow (e.g., \citealt{lazzati2018,troja2018b}). Admittedly, the prompt radiative efficiency (and therefore $E_{\rm j}$) can vary by a factor $\sim 2$, which would require a change in the external density and microphysical parameters by a similar factor. However, these changes will not affect the slope of the rise in the afterglow. We find that the observed frequency lies between the minimum and the cooling frequencies, in agreement with observations (\citealt{margutti2018}).

\begin{figure}
\includegraphics[width=\columnwidth]{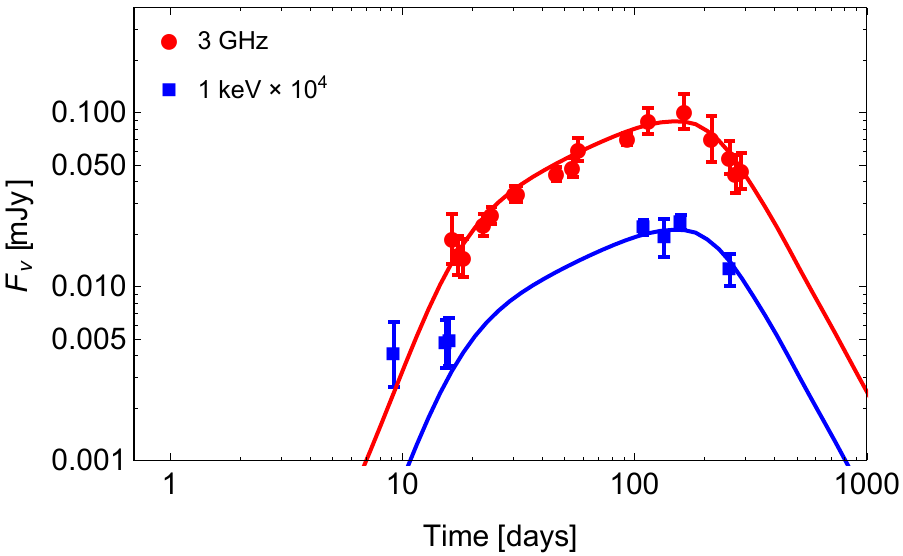}
\vspace*{-5mm}\caption{Afterglow light curves for our simulated structured jet (from Fig. \ref{jetstructure}) and observed data for comparison in radio (3 GHz, red points) and X-ray (1 keV, blue squares) (from \citealt{alexander2018,margutti2018}). Relevant parameters used are $E_{\rm j}\approx 5\times10^{50}$ erg, $n\approx 0.05$ cm$^{-3}$, $\epsilon_{\rm e}\approx 0.01$, $\epsilon_{\rm B}=10^{-4}$, $p=2.17$, $\theta_{\rm obs}=30^{\circ}$.}
    \label{afterglow}
\end{figure}
\section{discussion and conclusions}
\label{conclusions}
Determining how the power and Lorentz factor of a jet are distributed as a function of angle is a multi-scale endeavor. The simulations and analysis carried out in this work start at the central engine, where a jet is self consistently launched, and follows the jet out to larger distances as it collimates, accelerates and interacts with surrounding disk winds. The result is a more realistic description of the transverse jet structure that naturally produces an emission profile consistent with both the prompt and afterglow EM counterparts of GW170817. One aspect not considered here is interactions with the post-merger and dynamical ejecta that can take place at larger scales: in some extreme cases, this may even choke the jet and prevent it from breaking out (e.g., \citealt{gottlieb2018a}). However, it is likely that jet interactions with the ejecta will result in more energetic outflows at wider angles 
(e.g., \citealt{nagakura2014,duffell2015,berthier2017,bromberg2018,duffell2018}), that can potentially brighten the emission for off-axis observers at earlier times. This could be a reason why the first X-ray point at $\sim 10$ days in Fig. \ref{afterglow} is brighter than the afterglow model of our structured jet.

As seen in Fig. \ref{jetstructure}, the transverse jet structure in Lorentz factor remains approximately constant over most of the jet core before sharply declining at larger angles. In contrast, the energy per solid angle shows a much shallower decline for the majority of the jet's transverse extent. However, $\sim 90\%$ of the jet's energy is concentrated within a polar angle of $10^{\circ}$, indicating a narrow, energetic jet core in agreement with afterglow fits and the observations (e.g., \citealt{margutti2018,mooley2018,troja2018b}). 
The prompt and afterglow emission calculated using this structure obtained from our simulations match the observed data of GRB170817A well. The prompt emission profile in Fig. \ref{promptprofile} shows that for an event within the LIGO detectability volume, the count rate due to the prompt emission at an observing angle of $\sim 25^{\circ}$ is the order of $\sim 10^{2}-10^{3}$ counts/s (Sec.~\ref{prompt}). These values match the data obtained and inferred from GRB170817A (e.g., \citealt{abbott2017b,finstad2018}). In comparison, producing a similar result for the prompt emission with a top-hat jet would require $\theta_j\gtrsim 20^{\circ}$ (which is also the case for the afterglow; \citealt{lazzati2018}), this is not supported by observations of GRB170817A. As mentioned previously (e.g., \citealt{salafia2015,kathir2017}), the prompt emission profile for a structured jet has a much shallower drop off for larger observing angles compared to the emission profile of a uniform top-hat jet. Tied with a coincident LIGO trigger, this enables the detection of the prompt emission for substantially misaligned observers, making this signal a more feasible EM counterpart than previously thought. Indeed, Fig.~\ref{promptprofile} indicates the prompt emission would be detectable, with the aid of a coincident LIGO trigger, up to an observing angle of $\sim 20^{\circ}$ off-axis for a source at the edge of the LIGO detectability volume ($\sim$ 200 Mpc), and up to $\sim 30^{\circ}$ for events like GW170817 that are much closer. Let us assume the prompt emission of all short GRBs, from GW events detected by LIGO, are detectable up to a viewing angle of $20^{\circ}$. Then by integrating the detection probability of GW events (\citealt{schutz2011}) from an inclination angle of $0^{\circ}$ to $20^{\circ}$, we find that a fraction of $\sim 0.2$ GW events (out of all that produce short GRBs) will have a detectable prompt emission. However  this fraction could change appreciably for different jet structures (e.g., \citealt{beniamini2018}).

The afterglow light curves from our structured jet reproduce the observed rise, peak and decline of GRB170817A. The core of the jet begins to spread after it decelerates, which will steepen the decline in the afterglow light curve after the peak (for off-axis observers). The slope of the decline in our afterglow is $\approx -2.4$, in agreement with afterglow models where jet spreading is taken into account (e.g., \citealt{troja2018b}). From the afterglow modelling of our structured jet, we find that the temporal dependence of the observed rise strongly constrains the viewing angle to be close to 30$^{\circ}$, since larger (smaller) viewing angles will produce a steeper (shallower) rise in the afterglow.

\section*{Acknowledgements}
We thank the anonymous referee and Brian Metzger for providing helpful feedback. This material is based upon work supported by the National Science Foundation under Grants 1816694, 1815304 and 1816136. AK and DG acknowledge support from NASA grants NNX16AB32G and NNX17AG21G. AT acknowledges support from Northwestern University. RBD acknowledges support from the Chein Hu Endowment at the Department of Physics and Astronomy at Sacramento State. This research used resources of the National Energy Research Scientific Computing Center (NERSC), which is supported by the Office of Science of the U.S. Department of Energy under Contract No. DE-AC02-05CH11231. Computations were performed at Edison (repositories m1186, m2058, m2401, and the scavenger queue).

%%%%%%%%%%%%%%%%%%%%%%%%%%%%%%%%%%%%%%%%%%%%%%%%%%



%\bibliographystyle{mnras}

\bibliographystyle{mnras}
\bibliography{references}
%\input{mnras_template.bbl}

%%%%%%%%%%%%%%%%%%%%%%%%%%%%%%%%%%%%%%%%%%%%%%%%%%

%%%%%%%%%%%%%%%%% APPENDICES %%%%%%%%%%%%%%%%%%%%%

%%%%%%%%%%%%%%%%%%%%%%%%%%%%%%%%%%%%%%%%%%%%%%%%%%

% Don't change these lines
%\bsp	% typesetting comment
\label{lastpage}
\end{document}